\documentclass{vgtc}                          % final (conference style)
\ifpdf%                                % if we use pdflatex
  \pdfoutput=1\relax                   % create PDFs from pdfLaTeX
  \usepackage{graphicx}                % allow us to embed graphics files
  \DeclareGraphicsExtensions{.pdf,.png,.jpg,.jpeg} % for pdflatex we expect .pdf, .png, or .jpg files
\else%                                 % else we use pure latex
  \usepackage{graphicx}                % allow us to embed graphics files
  \DeclareGraphicsExtensions{.eps}     % for pure latex we expect eps files
\fi%

\graphicspath{{figures/}{pictures/}{images/}{./}} % where to search for the images

\usepackage{microtype}                 % use micro-typography (slightly more compact, better to read)
\PassOptionsToPackage{warn}{textcomp}  % to address font issues with \textrightarrow
\usepackage{textcomp}                  % use better special symbols
\usepackage{mathptmx}                  % use matching math font
\usepackage{times}                     % we use Times as the main font
\usepackage{cite}                      % needed to automatically sort the references
\usepackage{tabu}                      % only used for the table example
\usepackage{booktabs}                  % only used for the table example
\onlineid{0}

\vgtccategory{Research}

\vgtcinsertpkg

\title{Word Clouds in the Wild}

\author{Rebecca M.~M.~Hicke\thanks{e-mail: rmh327@cornell.edu}\\
\scriptsize Cornell University
\and Maanya Goenka\thanks{e-mail: goenkam@carleton.edu}\\
\scriptsize Carleton College
\and Eric Alexander\thanks{e-mail: ealexander@carleton.edu}\\
\scriptsize Carleton College} %

 \teaser{
   \centering
   \includegraphics[width=\linewidth]{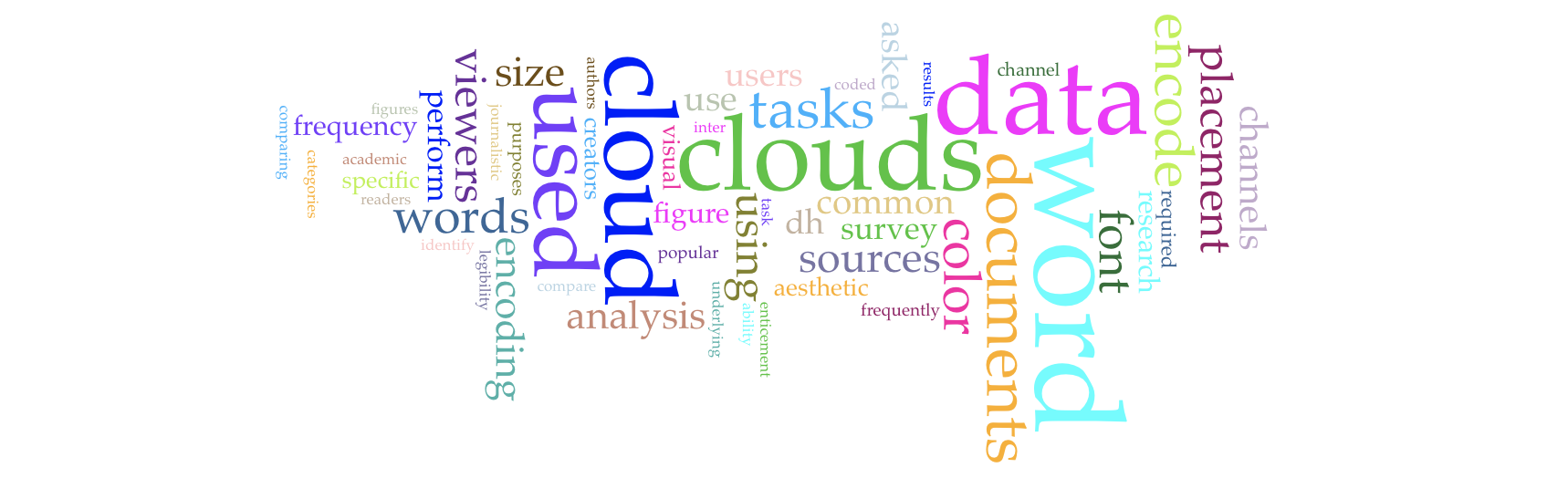}
   \caption{A word cloud of this paper, created with Voyant.}
   \label{fig:teaser}
 }

\abstract{
Word clouds are frequently used to analyze and communicate text data in many domains. In order to help guide research on improving the legibility of word clouds, we have conducted a survey of their usage in Digital Humanities academia and journalism. Using a modified grounded theory approach, we sought to identify the most common purposes for which word clouds were employed and the most common visual encodings they contained. Our findings indicate that font size, color, and word placement dominate as the primary data-encoding channels, as we hypothesized. Perhaps more surprisingly, we found that asking viewers to perform analytical tasks with word clouds was relatively common, especially in DH sources. This suggests that research into the interactions of these visual encoding channels (particularly in regards to legibility) is warranted.

} % end of abstract

\CCScatlist{Task Abstractions \& Application Domains; Social Science, Education, Humanities, Journalism, Intelligence Analysis, Knowledge Work; Text/Document Data
}

\begin{document}


\firstsection{Introduction}

\maketitle

Word clouds are frequently used to analyze and communicate text data in many domains. Particularly within the digital humanities, a field that requires considerable textual analysis, word clouds are a common method of presenting and analyzing data \cite{janicke2017visual, florian2018visualization}. Their intuitive and engaging design has made them a popular choice for authors, but many visualization specialists are skeptical about their effectiveness---a skepticism that has been reinforced by the results of several studies \cite{rivadeneira2007getting, hearst2019evaluation, felix2017taking}. To get a better understanding of which visual encodings are popular, and thereby help direct future research on word clouds' legibility and create a set of guidelines for proper word cloud usage, we conducted a survey on how word clouds are being used ``in the wild.''

Using a modified grounded theory approach, we began collecting data on several variables that previous research had indicated were important. We then added new codes to the dataset as we encountered interesting trends and figure attributes. We designed taxonomies to describe the possible values for several of these variables, which contributed to consistent and analyzable coding. Throughout the data collection process, we focused especially on two things:
\begin{itemize}
    \item the \textbf{tasks} that word cloud designers were asking their readers to perform
    \item the \textbf{design choices} that word cloud designers were making in determining how to encode their data.
\end{itemize}
In gathering this data, we hope to be able to assess whether common design choices are actually supporting the intended uses of the clouds. There are a variety of past experiments that indicate how effective word clouds are at affording different types of analysis. 
Studies have shown, for instance, that word clouds do not support the lookup and retention of specific words~\cite{rivadeneira2007getting}. 
Others have showed that differences in word shape can bias interpretation of font size~\cite{alexander2017perceptual}.
On the other hand, other experiments indicate that word clouds are effective at conveying a general ``gist'' of their contents in certain situations~\cite{alexander2016assessing}. More generally, it seems that the the effectiveness of word clouds depends dramatically on the task for which they are being used~\cite{felix2017taking}.
However, to our knowledge, there has not been a thorough analysis of how commonly readers are actually asked to perform these various tasks in real-world settings. 
We hope to try and match some of the above results against word cloud use in the wild, and potentially identify gaps in the research.

We have collected and analyzed over 500 word clouds across more than 300 documents found within academic literature in the field of Digital Humanities (DH) and journalistic sources such as the New York Times. We have focused initially on DH and journalism as two of the most prominent arenas in which word clouds appear, but hope to broaden our search going forward. Along with our analysis and proposal of future research questions, we provide this dataset of word cloud instances as an artifact for potential further study.

%\section{Related Works}
%\input{sections/relatedWorks}

\section{Methodology}
We used a modified version of grounded theory to direct our data collection process, which aimed to document instances of word clouds and word-cloud-adjacent figures in sources from DH academia and journalism. We searched for figures by hand in five academic DH sources, selected for their availability across multiple years and the accessibility of their materials: Digital Humanities Conference abstracts (2016-2018), the Journal of Digital Humanities (2011-2014), Digital Humanities Quarterly (2007-2020), The Journal of Interactive Technology and Pedagogy (2012 - 2020; although not specifically a DH journal, uses similar techniques and frequently addresses DH pedagogy), and Digital Scholarship in the Humanities (1996-2020). Within these sources, we identified 194 relevant figures in 120 documents. We then used APIs to query three journalistic sources – The New York Times, the Guardian, and Die Zeit – searching for common phrases used to describe word clouds (e.g. `word cloud,' `topic cloud,' `wordle,' `word visualization,' and `wortwolke,` in the case of the German Die Zeit). We found 351 relevant figures in 248 documents.

We coded each figure as belonging to one of three categories based on how closely we felt it resembled a word cloud. The three categories were: \emph{definitely} (word clouds), \emph{maybe} (word clouds), and \emph{edge cases}. 
%(See the \href{https://cs.carleton.edu/faculty/ealexander/wcitw}{supplemental materials} for details on terminology usage.) 
Clouds labeled as \emph{definitely} or \emph{maybe} word clouds were included in our analysis for this paper, but \emph{edge cases} were excluded.

We began our coding process by collecting data on a small set of variables that we hypothesized would be informative based on previous research. In particular, these variables included references to visual encodings that are often used and a handful of specific tasks \cite{alexander2016assessing, rivadeneira2007getting, bateman2008seeing, hearst2008tag, felix2017taking}. As coding continued, we added new fields to the dataset when we identified interesting trends and figure properties.
When we added a new field, we would retroactively code it for all past clouds. 
Some of the more complex or abstract fields required us to develop hierarchical taxonomies articulating their possible values (described in detail in the following section) to ensure they were coded in a regular and analyzable manner.
We ultimately collected data on 40 different aspects of the word clouds (and their containing documents). While we will only address the most salient results in this paper, the complete dataset will be made available as supplemental materials.
Our coding process began with an initial period of norming, in which two researchers independently coded 65 figures from DH academic sources and then met to resolve conflicts. After the norming was complete, the two primary coders worked on data collection independently but met weekly to get input on difficult judgements and compare notes.

\section{Taxonomies}
For three of the more complex/subjective fields we coded, we developed hierarchical taxonomies of the values we encountered, so as to afford easier comparison in analysis.

\subsection{Purpose of cloud}
\label{sec_purposeTaxonomy}

We tried our best to identify the rhetorical purpose of the clouds we encountered---that is, what the author hoped to accomplish by including them, or how they intended the reader to use them. This was difficult to ascertain for many of the clouds, especially those for which there was little description within the surrounding text. However, we were able to consolidate the uses we saw into three main categories, along with several sub-categories:

\begin{itemize}
    \item \textbf{Analysis:} Creators were using a word cloud to learn something more about the underlying data.
    \begin{itemize}
        \item \textbf{Exploration:} Creators were using the cloud to discover information about the data, but they did not have a specific goal about what they were hoping to find.
        \item \textbf{Hypothesis Generation:} Creators were attempting to form an explicit theory about a source from a cloud.
        \item \textbf{Hypothesis Testing:} Creators were using a cloud to confirm or refute a previously articulated thesis.
    \end{itemize}
    \item \textbf{Presentation:} Creators were using a word cloud to present information to viewers but were not expecting the viewers to form new conclusions about the underlying textual data. 
    \begin{itemize}
        \item \textbf{Providing Data:} Creators were using a cloud to efficiently summarize large amounts of data for the audience. While the creators may have been analyzing the cloud’s underlying data, they were not using the word cloud as an analysis tool itself.
        \item \textbf{Example of Form:} Creators were discussing visualizations in general or word clouds specifically and included a figure as an example of what a word cloud was.
        \item \textbf{Navigation:} Creators were using a cloud to show viewers what data was contained in a tool, website, or database and to allow them to search and filter that data.
    \end{itemize}
    \item \textbf{Enticement:} Creators were using a cloud to bring attention to the document it was found in. While the underlying textual data of the cloud may have been relevant to the topic of the document, viewers were not expected to seriously examine it. (If a cloud was used for multiple reasons, other goals superseded that of enticement in our labeling process.)
\end{itemize}

\subsection{Lower-level tasks}
\label{sec_tasksTaxonomy}

We also sought to identify the specific analysis tasks that it seemed authors intended readers to perform in their interactions with the clouds. We split these tasks into two categories based on how many clouds readers were being directed to compare:

\begin{itemize}
    \item \textbf{Intra-cloud comparison tasks:} These tasks required readers to examine data encoded within a single word cloud.
    \begin{itemize}
        \item \textbf{Browsing:} Viewers were asked to use the figure to navigate a tool or to filter information in a website or database.
        \item \textbf{Gist-forming:} Viewers were asked to extract the central idea or thesis of the underlying textual data in the cloud. 
        \item \textbf{Summarization:} Viewers were asked to extract a general summary of the cloud’s underlying textual data, which could include more than its primary focus. 
        \item \textbf{Finding ‘topics’, groups, or clusters:} Viewers were asked to find groups of semantically associated words within a cloud.
        \item \textbf{Finding the biggest word(s):} Viewers were asked to identify the largest word or words in a cloud.
        \item \textbf{Comparing specific words:} Viewers were asked to compare the prominence of two or more specific words in a cloud.
        \item \textbf{Searching for specific word(s):} Viewers were asked to find one or more specific words in a cloud.
    \end{itemize}
    \item \textbf{Inter-cloud comparison tasks:} These tasks required readers to compare data encoded across \emph{multiple} clouds.
    \begin{itemize}
        \item \textbf{Comparing specific word(s):} Viewers were asked to compare the frequency of a specific word or words between clouds.
        \item \textbf{Finding and comparing biggest word(s):} Viewers were asked to identify and then compare the largest word or words in multiple clouds.
        \item \textbf{Comparing ‘topics’ or clusters:} Viewers were asked to compare which groups of semantically associated words appeared in multiple clouds.
        \item \textbf{Comparing gists:} Viewers were asked to compare the gists, or central theses, of multiple clouds.
    \end{itemize}
\end{itemize}

\subsection{Visual encodings}
\label{sec_encodingsTaxonomy}

The final taxonomy we created describes how different visual variables present within a word cloud (font size, font weight, color, placement, direction, and font) were used to encode data (or not). We saw six main uses for these visual channels:
\begin{itemize}
    \item \textbf{Quantitative:} The channel is used to encode quantitative data.
    \item \textbf{Qualitative:} The channel is used to encode qualitative data.
    \item \textbf{Alphabetical:} The channel is used to encode an alphabetical ordering.
    \item \textbf{Highlighting:} The channel is used to highlight particular elements/terms.
    \item \textbf{Aesthetics:} The channel is used purely for aesthetic purposes.
    \item \textbf{Consistent:} The channel is not being used in this figure; this variable does not vary. (E.g., all words in the figure are the same color.)
\end{itemize}

\section{Results}
The primary trends we identified within the data can be divided into
three categories:
how visual encoding channels were used, what tasks users were asked to perform with word clouds, and what tools were used to create word clouds. 

\subsection{Visual encoding channels}
\label{sec:channelsResults}

The visual channels present in a word cloud (font size, font weight, color, placement, direction, and font) can be used to encode data, enhance the cloud's aesthetic appeal, or can be held consistent. The channel most frequently used to encode data was font size, which was used to encode data, almost always quantitative data, in over 92\% of documents. 
The next most popular channel for encoding data was placement (44\% of documents)---again, it was used primarily for quantitative data (35\%) but also occasionally for qualitative data (7\%) or alphabetical ordering (6\%).
The data placement most commonly encoded was word frequency (77\% of documents). This was often used redundantly with font size; when font size encoded word frequency, placement was used redundantly in 31\% of documents.

\begin{figure}[!ht]
    \centering
    \includegraphics[scale=0.5]{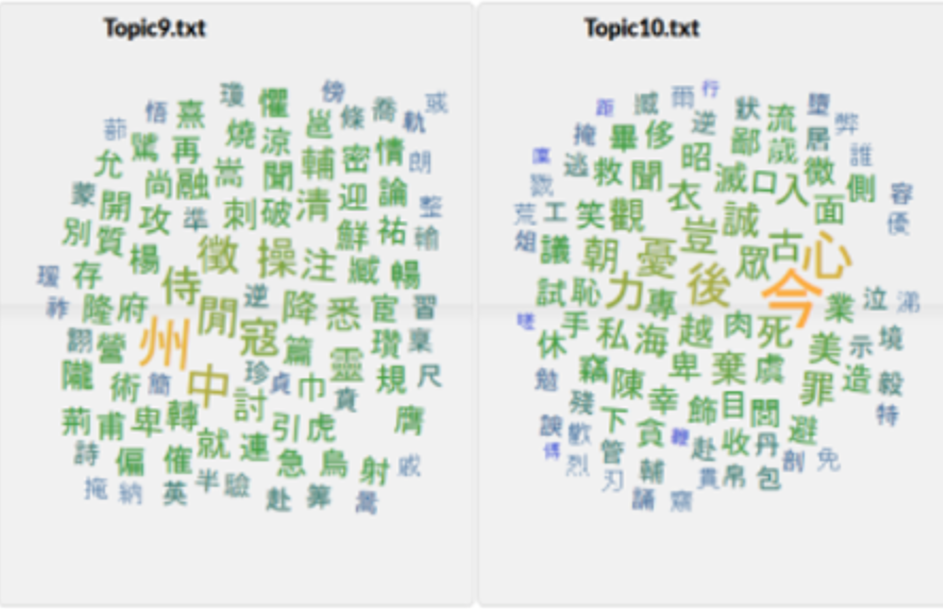} \\
    \caption{This figure uses color, placement, and font size redundantly to encode frequency (from “Lexos: An Integrated Lexomics Workflow” by S.~Kleinman in the 2017 DH Conference abstracts).}
    \label{fig:threeRedundant}
\end{figure}

The third most popular data-encoding channel was color (18\% of documents). Color was used more frequently to encode quantitative data (11\% of documents) than to encode qualitative data (10\%) or highlight words (2\%). However, in the documents in which color was used to encode data, it was used to encode frequency only 46\% of the time---a much lower proportion than font size and placement. Additionally, only around 9\% of documents that used font size to encode frequency used color redundantly. Even less common was the use of font size, placement, and color all encoding frequency together; this only occurred in around 2\% of documents. An example of this three-fold redundant encoding can be seen in Fig. \ref{fig:threeRedundant}.

\begin{figure}[!ht]
    \centering
    \includegraphics[scale=0.23]{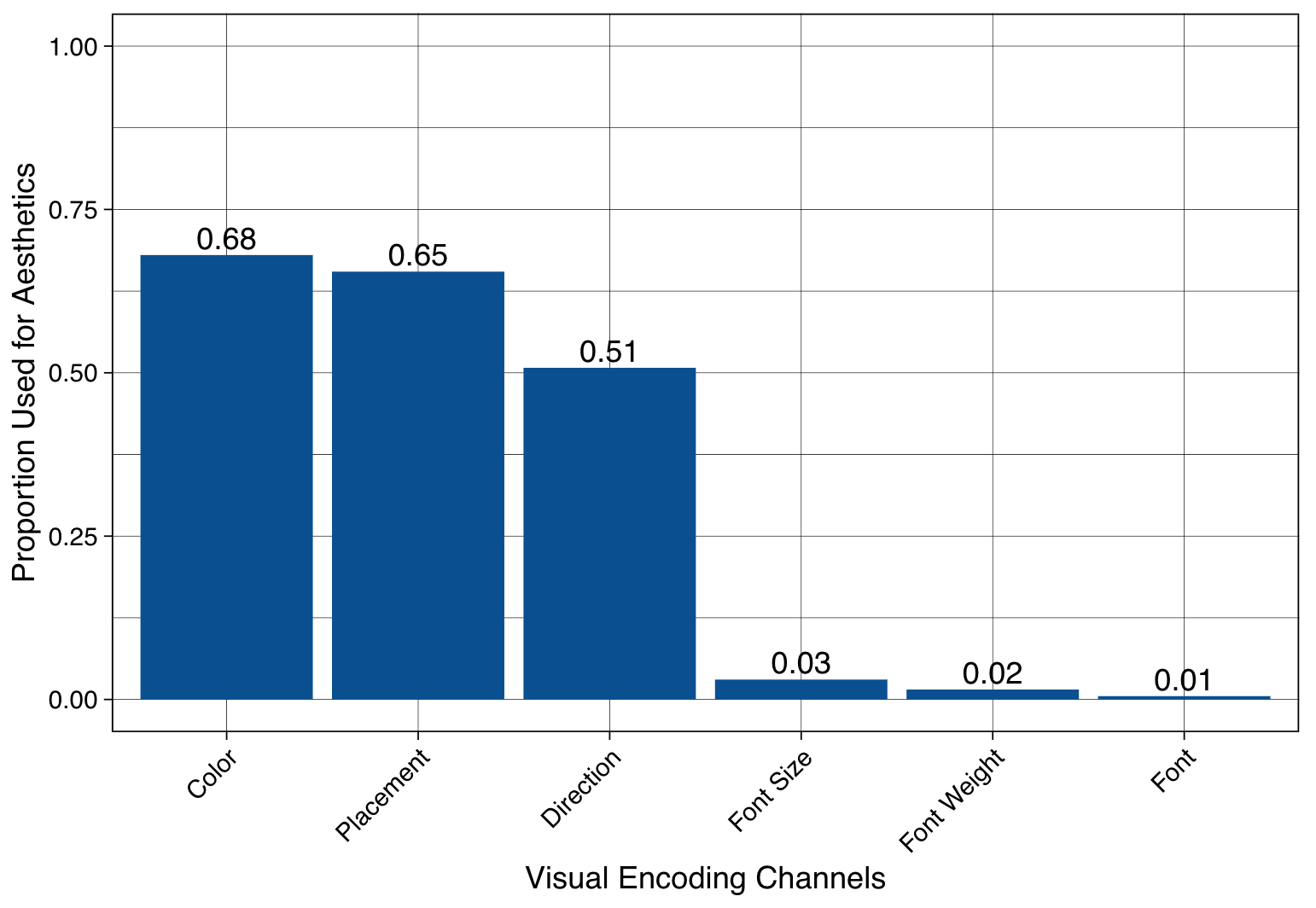} \\
    \caption{Proportion of documents in which channels are used for aesthetics}
    \label{fig:propAesthetics}
\end{figure}
\begin{figure}[!ht]
    \centering
    \includegraphics[scale=0.23]{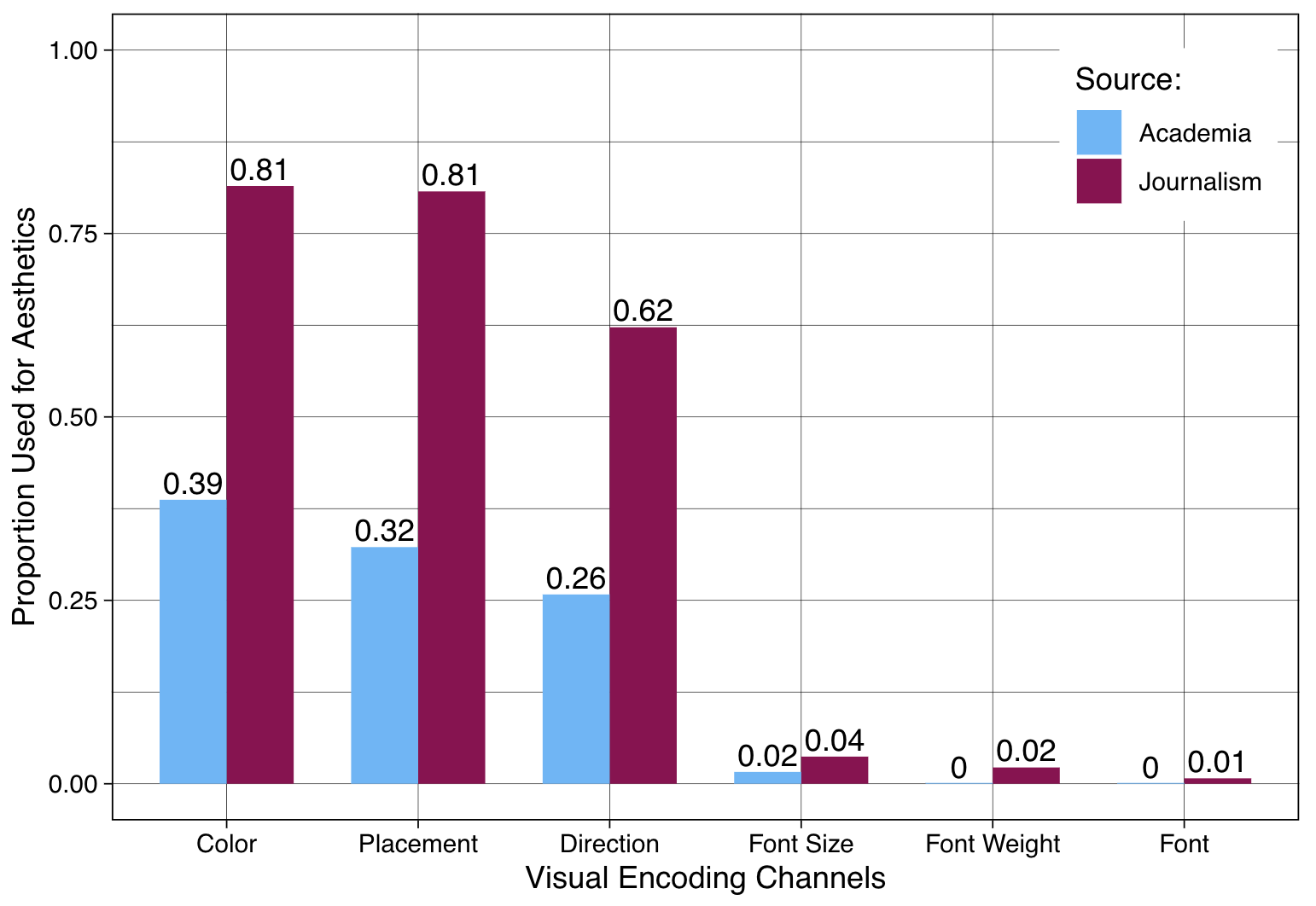} \\
    \caption{Proportion of documents in which channels are used for aesthetics – Journalism vs.~Academia}
    \label{fig:propAestheticsJvA}
\end{figure}

The color, placement, and direction of words were each used for aesthetic purposes in over half of the documents we coded (Fig. \ref{fig:propAesthetics}). Each of these three encoding channels were used far more commonly for aesthetic purposes in journalistic sources, but even in DH academic sources they were used for aesthetics in over 25\% of documents (Fig. \ref{fig:propAestheticsJvA}).

\subsection{Tasks}
\label{sec:tasksResults}

Authors most frequently intended their word clouds to be used for providing data (50\% of documents), analytic exploration (37\% of documents), or enticement (22\% of documents). We found that enticement was primarily popular in journalistic sources and appeared in less than 2\% of the documents from DH academic sources.

\begin{figure}[!ht]
    \centering
    \includegraphics[scale=0.25]{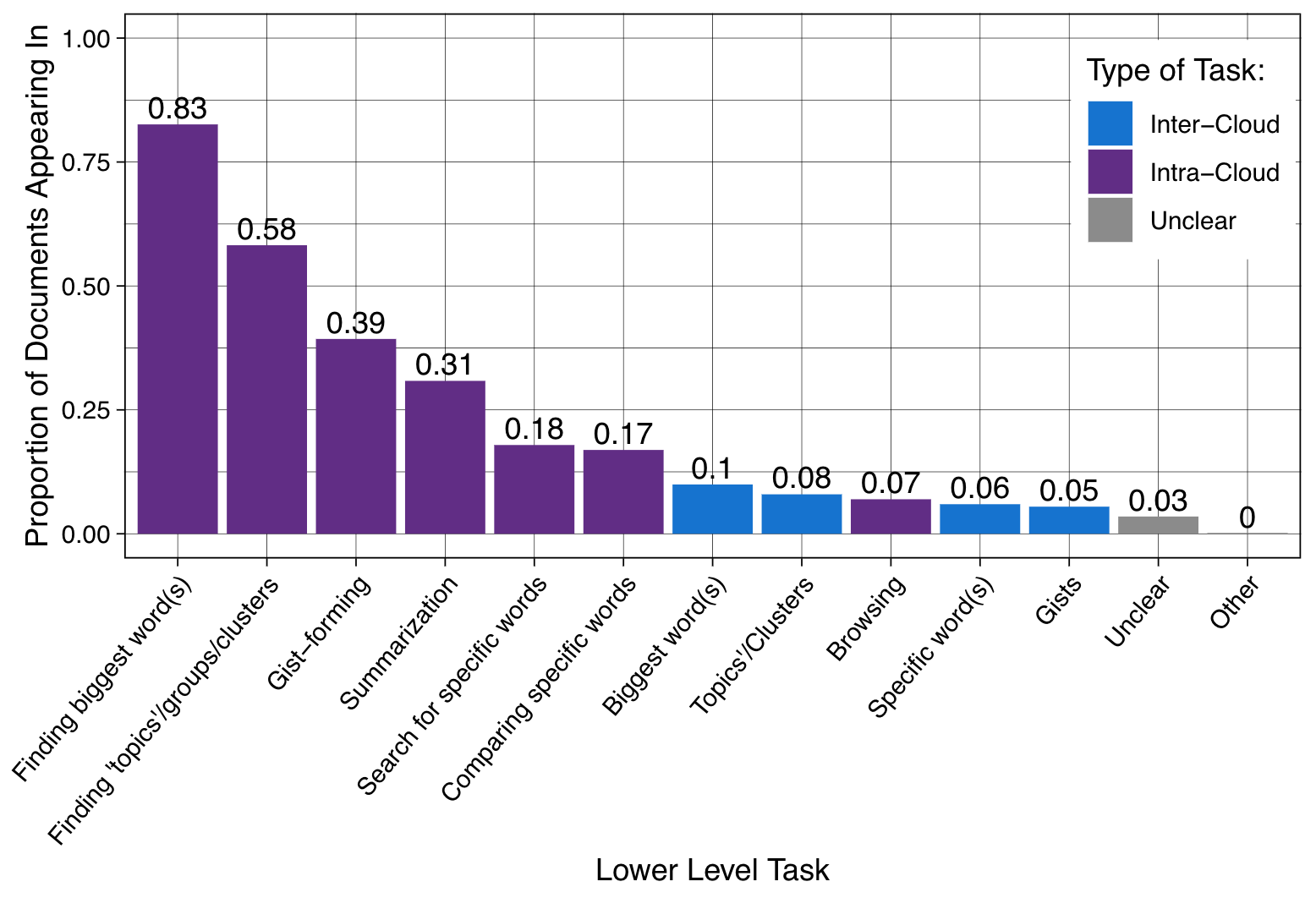} \\
    \caption{Proportion of total documents that require users to perform lower level tasks}
    \label{fig:propTasks}
\end{figure}

Users were required to perform both intra- and inter-cloud analysis tasks to interpret the word clouds we found (Fig. \ref{fig:propTasks}). All of the documents coded required users to perform intra-cloud analysis and about 1/5 
of the documents required users to perform inter-cloud analysis. Interestingly, DH academic sources asked for inter-cloud comparisons more frequently than journalistic ones (32\% vs 15\% of documents).

\begin{figure}[!ht]
    \centering
    \includegraphics[scale=0.3]{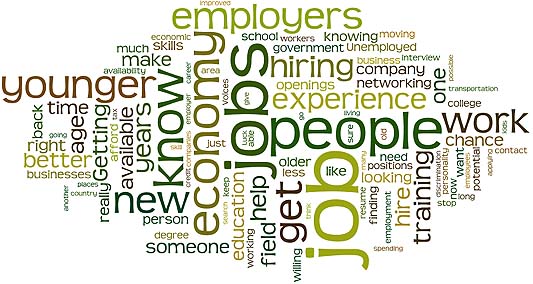} \\
    %\: \\
    {\large “Here’s a word cloud of responses showing how prevalent age-related answers were.”}
    \\
    \caption{Viewers are asked to find semantic groupings of words in this cloud (from “Unemployed, and Wanting to Be Younger” by C.~Rampell in the New York Times blog Economix: Explaining the Science of Everyday Life).}
    \label{fig:findTopics}
\end{figure}

Our survey found that the most common lower-level intra-cloud analysis task required of users was finding the biggest word(s) in a cloud. Another commonly required task was asking users to find semantic groupings of words within a cloud (58\% of documents). Fig. \ref{fig:findTopics} provides an example of a cloud that requires users to perform this task. Gist-forming and summarization were also frequently required of users (39\% and 31\% of documents respectively).

\subsection{Tools}
\label{sec:channelsResults}

\begin{figure}[!ht]
    \centering
    \includegraphics[scale=0.25]{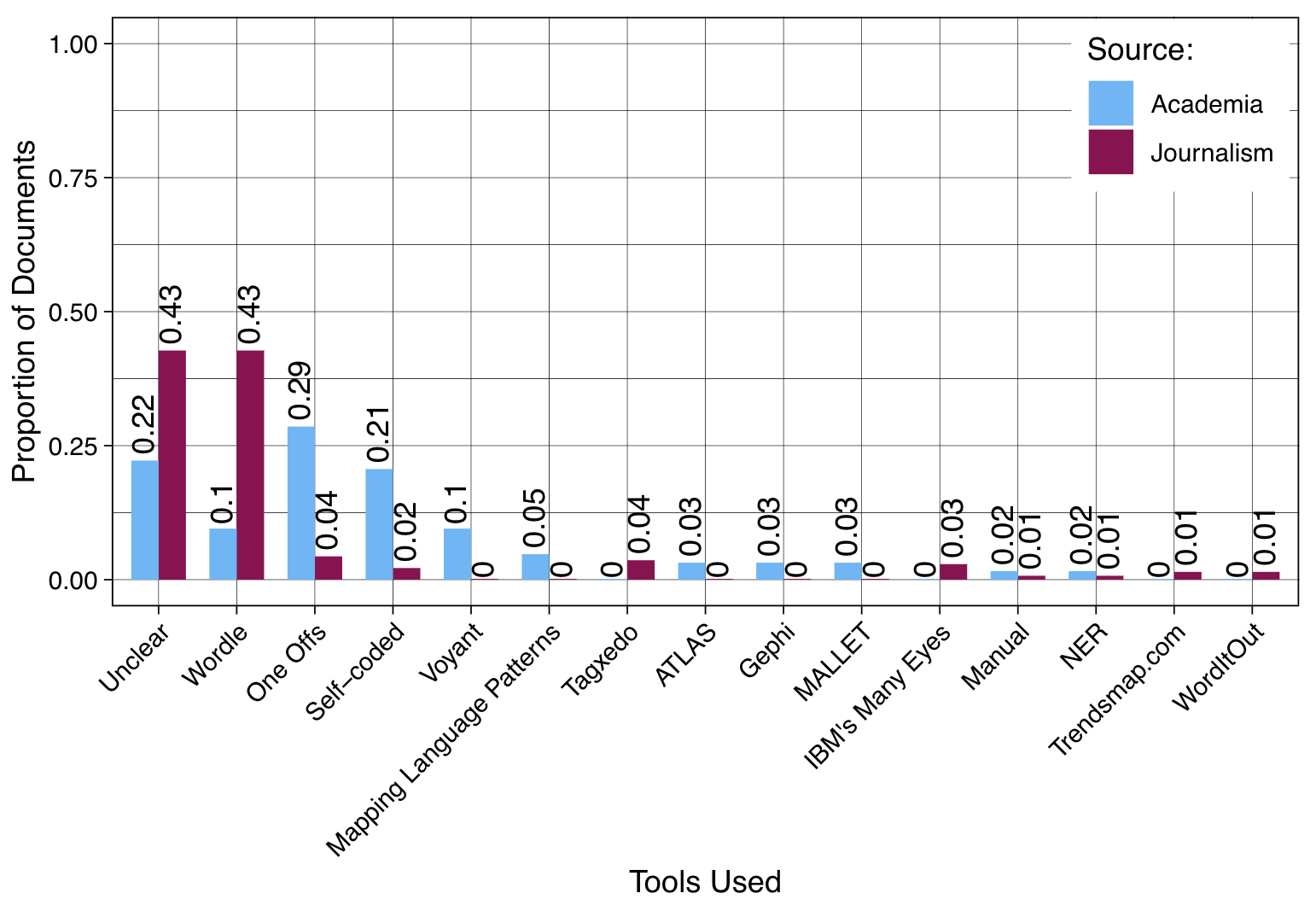} \\
    \caption{Tools used to create clouds – Journalism vs.~Academia}
    \label{fig:propTools}
\end{figure}

Many of the clouds we coded were created using pre-built tools and around a third of the documents contained clouds made with Wordle specifically. Just over a third of the documents did not report the method used to create at least one of their clouds. It was more common for authors to use Wordle or to not report how a cloud was created in journalistic sources, whereas unique tools and self-coded clouds appeared more frequently in the DH sources (Fig. \ref{fig:propTools}).

\section{Discussion}
The results of our survey point to several fruitful avenues for future research into improving the legibility of word clouds.

\subsection{Visual encoding channels}
\label{sec:channelsDiscussion}

\begin{figure}[!ht]
    \centering
    \includegraphics[scale=0.5]{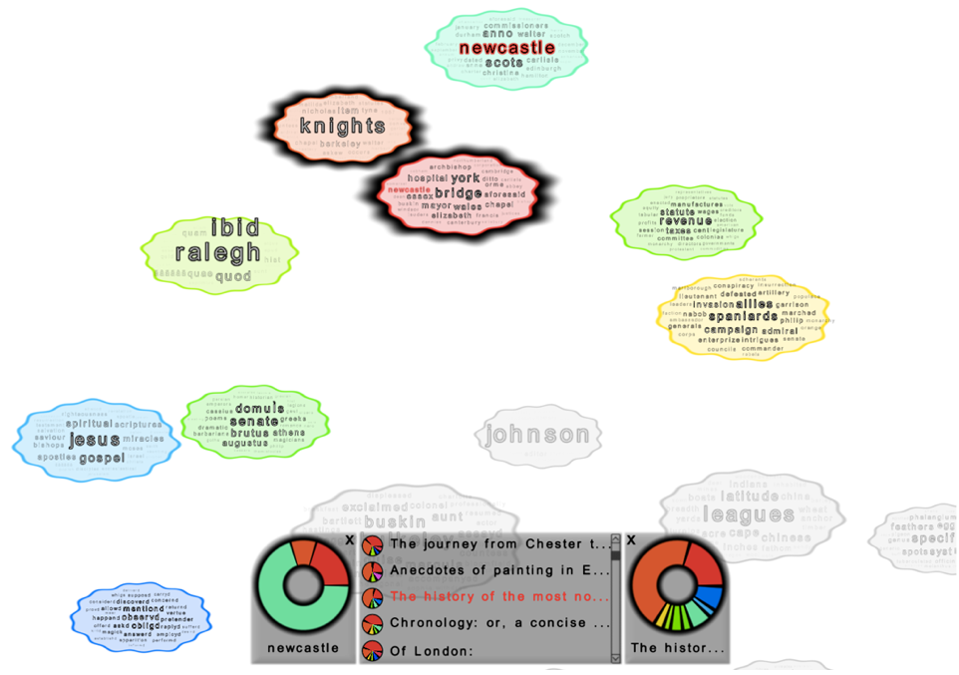} \\
    \caption{This figure uses an unusually large variety of visual encodings. The color and placement of both words and clouds are encoding a variety of data (from “Exploratory Search Through Visual Analysis of Topic Models” by P.~Jänichen et al in DHQ 11.2).}
    \label{fig:manyVisualEncodings}
\end{figure}

The survey revealed several trends in how word cloud creators used common visual encoding channels. The vast majority of the figures that we identified as word clouds used font size to encode data. The fact that so many figures integrated this aspect of a ‘typical’ or ‘traditional’ word cloud suggests that this may be an identifying characteristic of word clouds that should be maintained if we wish figures to be recognizable as such.

Additionally, because placement was such a popular choice for encoding data, especially frequency data, it is worth studying how its use impacts users’ understanding of a cloud. We are particularly interested in investigating the redundant use of placement, color, and font size and whether this helps or harms a cloud's legibility---especially when contrasted with the aesthetic or random variance in placement and color that exists in many clouds.
In particular, it seems pertinent to study how the random placement of words in a cloud affects a user’s ability to identify semantic groupings of words. Hearst et al.~% in “An Evaluation of Semantically Grouped Word Cloud Designs” 
found that spatially grouping semantically related words in a cloud made topic recognition tasks easier \cite{hearst2019evaluation}. The next section of the discussion will show that word cloud designers frequently ask their viewers to identify topics in word clouds, yet placement is rarely used to encode qualitative data, meaning that authors in our survey followed the advice from Hearst et al.~infrequently at best. 

The results of the survey also reveal that color was used infrequently to encode data of any sort. This suggests that it may be available as a currently ‘unused’ channel with which to try and mitigate the common perceptual downsides of word clouds. In addition, the results show that, when it was encoding data, color was used to encode data other than frequency a large proportion of the time. This indicates that color can be used to encode a wide variety of data and suggests that researching whether there is a perceptual downside to using color and font size to encode different data would be valuable.

The infrequency with which authors used font size and color redundantly to encode frequency means that it is also worth examining whether using color and font size together would improve users’ ability to interpret frequency data. Research into whether the even rarer design choice – using color, font size, and placement together to encode frequency – may improve the legibility of clouds appears to be another valuable avenue for future research.

\begin{figure}[!ht]
    \centering
    \includegraphics[scale=0.65]{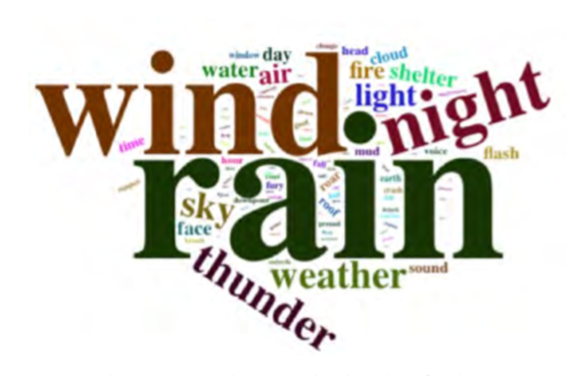}\includegraphics[scale=0.65]{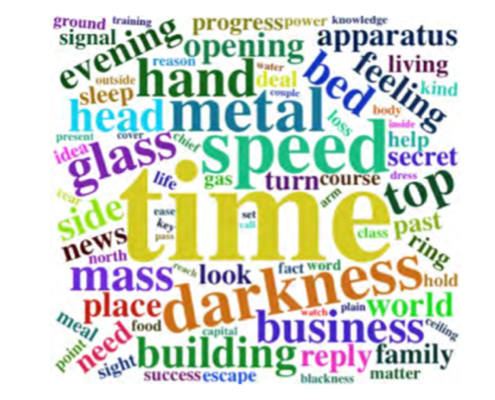} \\
    \caption{An example of color and direction used for purely aesthetic purposes in a DH source (from “SciFiQ and ‘Twinkle, Twinkle’: A Computational Approach to Creating ‘the Perfect Science Fiction Story’” by A.~Hammond and J.~Brooke in the 2018 DH abstracts).}
    \label{fig:colorDirectionAcademic}
\end{figure}

Ultimately, though we were curious to investigate the use of other visual variables such as word orientation, our survey indicates that the use of font size, color, and word placement dramatically overshadows the use of any other channels.
While it remains worth investigating how the use of other channels for aesthetic purposes (see Fig. \ref{fig:colorDirectionAcademic}) might harm the \emph{legibility} of a cloud, nothing in our data indicates that it would be advisable to use such other channels for data encoding. To do so may risk bucking convention past what would be identifiable as a word cloud to many readers.

The frequency with which direction, color, and placement were used for aesthetic purposes prompts two major questions that warrant further study. Is the aesthetic use of these channels affecting users’ ability to parse the relevant data? And could using these channels to help encode data instead, as suggested earlier in this section, augment users’ ability to understand word clouds? 

\subsection{Tasks}
\label{sec:tasksDiscussion}

\begin{figure}[!ht]
    \centering
    \includegraphics[scale=0.75]{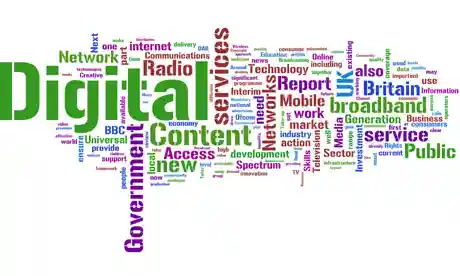} \\
    \caption{A cloud with mock data used for enticement in “Can design inspiration reduce the Guardian’s digital environmental impacts?” by S.~Wood in The Guardian.}
    \label{fig:mockData}
\end{figure}

Our survey found that authors intended word clouds to be used in a wide variety of ways and required users to perform many tasks in order to use clouds for those purposes. The survey also has implications as to the importance of being able to accurately interpret the data in word clouds. Enticement was identified as the third most common purpose for word clouds, and when word clouds are used for enticement the legibility of their underlying data may be less important. However, clouds were mostly used for enticement in journalistic sources. Readers were still frequently meant to interpret or extract data from word clouds, particularly in DH academic sources. Therefore, ensuring that aesthetic choices do not impact the legibility of data-encoding channels is important. The two most popular word cloud purposes, providing data and analytic exploration, focus on the general exploration of a cloud’s underlying data. Thus, facilitating a large number of lower-level tasks that allow for the highly accurate interpretation of various aspects of a dataset is valuable.

\begin{figure}[!ht]
    \centering
    \includegraphics[scale=0.7]{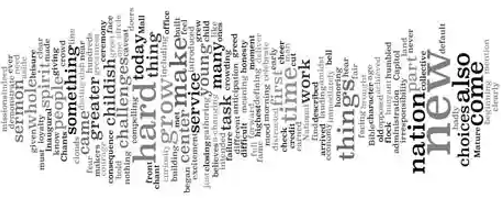}\\
    \includegraphics[scale=0.7]{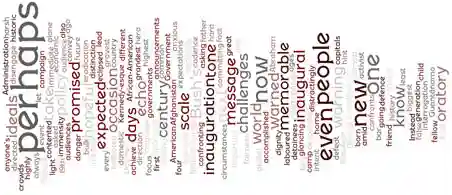} \\
    %\: \\
    {\large “But what about turning the tables on the commentators, and seeing how their cloudmaps compare to their politically different colleagues?”} \\
    \caption{An example of two clouds that require users to perform inter-cloud comparisons; the quote provides evidence (from “Wordsmiths or just wordy?” by H.~Siddique in The Guardian).}
    \label{fig:interCloud}
\end{figure}

It is surprising that such a high proportion of documents required inter-cloud analysis given the seeming perceptual difficulty of such tasks. The popularity of these tasks means that visualization designers should carefully research how to successfully facilitate inter-cloud comparisons or discourage using word clouds for this purpose. 

A common intra-cloud analysis task users were asked to perform was looking for semantic groupings of words within a cloud. Identifying semantic groupings is perceptually difficult, and word cloud creators generally don't use placement or color to help facilitate this task (as suggested by \cite{hearst2019evaluation}). In fact, the frequency with which color is used for aesthetic purposes may be making this task more difficult. This seems like a clear place in which perceptual guidelines for the creation of word clouds may have an impact and prevent authors from choosing aesthetic encodings that actively disrupt users’ ability to perform the intended analysis tasks.

The popularity of the gist-forming and summarizations tasks also indicates that future research should address how accurately viewers can perform these tasks and what design choices may affect their ability to do so. It is interesting to note that tasks that focus on extracting a general impression of the data behind a cloud (such as summarization) are more popular in journalistic sources, whereas tasks that focus on specific data points (such as searching for specific words) are more common in DH sources.

\section{Conclusion}

In this paper, we have presented the results of a survey examining the current usage of word clouds in journalism and DH academia. The survey revealed that word clouds remain a popular data visualization among journalists and DH academics, although they are not as popular as they were in the mid-2010s. The results of this survey point to the need for future experiments that will aid in creating a set of guidelines that will allow authors to create legible and visually appealing word clouds. We have determined that it will be particularly valuable to study the usage of font size, color, and placement as data-encoding channels, including the benefits of using them for redundant encoding and the possible side effects of using them for aesthetic purposes or to encode different data values. In particular, the survey suggests it will be important to research how these design choices impact viewers’ abilities to perform common lower-level analysis tasks such as finding the biggest word(s) in a cloud, identifying semantic groups of words, and finding the gist or summary of a cloud's underlying data.

\acknowledgments{This work was supported in part by the Public Works Grant and the Towsley Fellowship at Carleton College.}

\bibliographystyle{abbrv-doi}

\bibliography{template}
\end{document}